\documentclass[runningheads]{llncs}

\pdfoutput=1

\usepackage{cite}
\usepackage{amsmath,amssymb,amsfonts}
\usepackage{algorithmic}
\usepackage{graphicx}
\usepackage{textcomp}
\usepackage[table]{xcolor}
\usepackage{caption}
\usepackage{subcaption}
\usepackage{listings}
\usepackage{xparse}
\usepackage{stfloats}
\usepackage{tabularx}
\usepackage{ragged2e}
\usepackage{array}
\usepackage{cprotect}
\usepackage{hhline}
\usepackage{fixltx2e}
\usepackage{booktabs}
\usepackage{multirow}
\usepackage{graphics}
\usepackage{tcolorbox}
\usepackage{float}
\usepackage{enumitem}
\usepackage{longtable}

\usepackage{tikz}
\usetikzlibrary{
  arrows.meta, arrows, chains, positioning, shapes.geometric, decorations.pathreplacing,
  shadows, shapes, calc, shapes.symbols, shapes.callouts,patterns
}
\tikzstyle{input} = [coordinate]
\tikzstyle{output} = [coordinate]
\tikzset{font=\tiny}

\makeatletter
\pgfdeclareshape{document}{
\inheritsavedanchors[from=rectangle] 
\inheritanchorborder[from=rectangle]
\inheritanchor[from=rectangle]{center}
\inheritanchor[from=rectangle]{north}
\inheritanchor[from=rectangle]{north east}
\inheritanchor[from=rectangle]{south}
\inheritanchor[from=rectangle]{west}
\inheritanchor[from=rectangle]{east}
\backgroundpath{
\southwest \pgf@xa=\pgf@x \pgf@ya=\pgf@y
\northeast \pgf@xb=\pgf@x \pgf@yb=\pgf@y
\pgf@xc=\pgf@xb \advance\pgf@xc by-7.5pt 
\pgf@yc=\pgf@yb \advance\pgf@yc by-7.5pt
\pgfpathmoveto{\pgfpoint{\pgf@xa}{\pgf@ya}}
\pgfpathlineto{\pgfpoint{\pgf@xa}{\pgf@yb}}
\pgfpathlineto{\pgfpoint{\pgf@xc}{\pgf@yb}}
\pgfpathlineto{\pgfpoint{\pgf@xb}{\pgf@yc}}
\pgfpathlineto{\pgfpoint{\pgf@xb}{\pgf@ya}}
\pgfpathclose
\pgfpathmoveto{\pgfpoint{\pgf@xc}{\pgf@yb}}
\pgfpathlineto{\pgfpoint{\pgf@xc}{\pgf@yc}}
\pgfpathlineto{\pgfpoint{\pgf@xb}{\pgf@yc}}
\pgfpathlineto{\pgfpoint{\pgf@xc}{\pgf@yc}}
}
}
\makeatother

\newlength{\twosubht}
\newsavebox{\twosubbox}

\newtcbox{\highlight}[0]{boxsep=0pt,left=0pt,top=0pt,bottom=0pt,right=0pt,boxrule=0pt,arc=0pt,auto outer arc,colback=green,width=6cm}

\makeatletter
\DeclareFontFamily{U}{tipa}{}
\DeclareFontShape{U}{tipa}{m}{n}{<->tipa10}{}
\newcommand{\arc@char}{{\usefont{U}{tipa}{m}{n}\symbol{62}}}%

\newcommand{\arc}[1]{\mathpalette\arc@arc{#1}}

\newcommand{\arc@arc}[2]{%
  \sbox0{$\m@th#1#2$}%
  \vbox{
    \hbox{\resizebox{\wd0}{\height}{\arc@char}}
    \nointerlineskip
    \box0
  }%
}
\makeatother

\newcolumntype{C}{>{\Centering\arraybackslash}X}

\newcommand\ie{\emph{i.e.}\ }
\newcommand\eg{\emph{e.g.}\ }

\NewDocumentCommand{\codeword}{v}{%
\texttt{\textcolor{black}{#1}}%
}
\NewDocumentCommand{\codewordbf}{v}{%
\textbf{\texttt{\textcolor{black}{#1}}}%
}
\def\BibTeX{{\rm B\kern-.05em{\sc i\kern-.025em b}\kern-.08em
    T\kern-.1667em\lower.7ex\hbox{E}\kern-.125emX}}
\begin{document}

\title{Automatically Building Diagrams for Olympiad Geometry Problems}

\author{
  Ryan Krueger\inst{1}\orcidID{0000-0001-6856-0248} \and
  Jesse Michael Han\inst{2} \and
  Daniel Selsam\inst{3}
}
\authorrunning{R. Krueger et al.}
%
\institute{
  University of Oxford, Oxford, UK \and
  University of Pittsburgh, Pittsburgh PA, USA \and
  Microsoft Research, Redmond WA, USA\\
}

\maketitle

\begin{abstract}
  We present a method for automatically building diagrams for olympiad-level geometry problems and implement our approach in a new open-source software tool, the Geometry Model Builder (GMB).
  Central to our method is a new domain-specific language, the Geometry Model-Building Language (GMBL), for specifying geometry problems along with additional metadata useful for building diagrams.
  A GMBL program specifies (1) how to parameterize geometric objects (or sets of geometric objects) and initialize these parameterized quantities, (2) which quantities to compute directly from other quantities, and (3) additional constraints to accumulate into a (differentiable) loss function.
  A GMBL program induces a (usually) tractable numerical optimization problem whose solutions correspond to diagrams of the original problem statement, and that we can solve reliably using gradient descent.
  Of the 39 geometry problems since 2000 appearing in the International Mathematical Olympiad, 36 can be expressed in our logic and our system can produce diagrams for 94\% of them on average.
  To the best of our knowledge, our method is the first in automated geometry diagram construction to generate models for such complex problems.

\end{abstract}

\begin{figure}[t]
    \centering
    \includegraphics[width=\columnwidth,keepaspectratio]{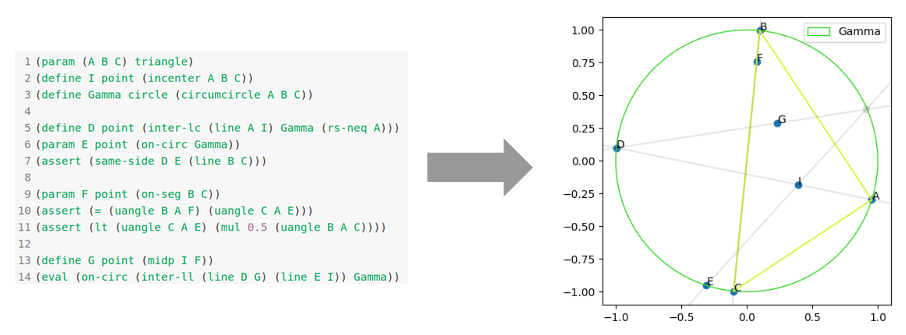}
    \caption{An example GMBL program and corresponding diagram generated by the GMB for IMO 2010 Problem 2.}
    \label{fig:ex}
\end{figure}

\section{Introduction}

Automated theorem provers for Euclidean geometry often use numerical models (\ie diagrams) for heuristic reasoning, \eg for conjecturing subgoals, pruning branches, checking non-degeneracy conditions, and selecting auxiliary constructions.
However, modern solvers rely on diagrams that are either supplied manually~\cite{gan2019automatically, wang2015automated} or generated automatically via methods that are severely limited in scope~\cite{itzhaky2013solving}.
Motivated by the IMO Grand Challenge, an ongoing effort to build an AI that can win a gold medal at the International Mathematical Olympiad (IMO), we present a method for expressing and solving olympiad-level systems of geometric constraints. 

Historically, algebraic methods are the most complete and performant for automated geometry diagram construction but suffer from degenerate solutions and, in the numerical case, non-convexity.
These methods are restricted to relatively simple geometric configurations as poor local minima arise via large numbers of parameters.
Moreover, degenerate solutions manifest as poor distributions for the vertices of geometric objects (\eg a non-sensical triangle) as well as intersections of objects at more than one point (\eg lines and circles, circles and circles).

We constructed a domain-specific language (DSL), the Geometry Model-Building Language (GMBL), to express geometry problems whose semantics induce tractable numerical optimization problems.
The GMBL includes a set of commands with which users introduce geometric objects and constraints between these objects.
There is a direct interpretation from these commands to the parameterization of geometric objects, the computation of geometric quantities from existing ones, and additional numerical constraints.
The GMBL employs \emph{root selector} declarations to disambiguate multiple solution problems, \emph{reparameterizations}
both to reduce the number of parameters and increase uniformity in model variance, and \emph{joint distributions} for geometric objects that are susceptible to degeneracy (\ie triangles and polygons).
Our DSL treats points, lines, and circles as first-class citizens, and the language can be easily extended to support additional high-level features
in terms of these primitives.

\begin{figure*}[t]
    \centering
    \includegraphics[width=\textwidth,height=\textheight,keepaspectratio]{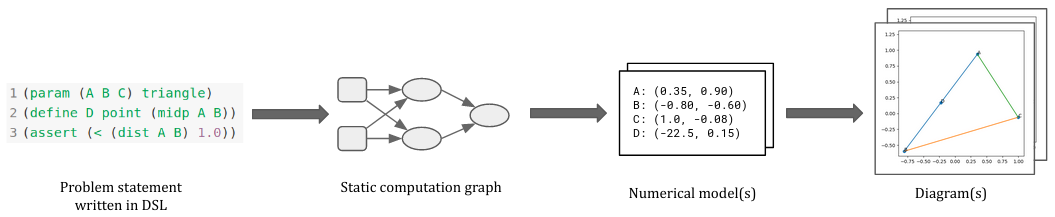}
    \caption{An overview of our method.
      Our program takes as input a GMBL program and translates it to a set of real-valued parameters and differentiable losses in the form of a static computation graph.
      We then apply gradient-based optimization to obtain numerical models and display them as diagrams.}
    \label{fig:overview}
\end{figure*}

We provide an implementation of our method, the Geometry Model Builder (GMB), that compiles GMBL programs into Tensorflow computation graphs~\cite{abadi2016tensorflow} and generates models via off-the-shelf, gradient-based optimization.
Figure \ref{fig:overview} demonstrates an overview of this implementation.
Experimentally, we find that the GMBL sufficiently reduces the parameter space and mitigates degeneracy to make our target geometry amenable to numerical optimization.
We tested our method on all IMO geometry problems since 2000 ($n=39$), of which 36 can be expressed as GMBL programs.
Using default parameters, the GMB finds a single model for 94\% of these 36 problems in an average of 27.07 seconds.
Of the problems for which our program found a model and the goal of the problem could be stated in our DSL, the goal held in the final model 86\% of the time. 

All code is available on GitHub\footnote{https://github.com/rkruegs123/geo-model-builder} with which users can write GMBL programs and generate diagrams.
Our program can be run both as a command-line tool for integration with theorem provers or as a locally-hosted web server.

\section{Background}

Here we provide an overview of olympiad-level geometry problem statements, as well as several challenges presented by the associated constraint problems.

\subsection{Olympiad-Level Geometry Problem Statements}

IMO geometry problems are stated as a sequential introduction of potentially-constrained geometric objects, as well as additional constraints between entities.
Such constraints can take one of two forms: (1) \emph{geometric} constraints describe the relative position of geometric entities (\eg two lines are parallel) while (2) \emph{dimensional} constraints enforce specific numerical values (\eg angle, radius).
Lastly, problems end with a goal (or set of goals) typically in the form of geometric or dimensional constraints.
The following is an example from IMO 2009:
\begin{equation}
  \tag{IMO 2009 P2}\label{eq:imo-2009-p2}
  \parbox{\dimexpr\linewidth-4em}{%
    \strut
    Let $ABC$ be a triangle with circumcentre $O$.
    The points $P$ and $Q$ are interior points of the sides $CA$ and $AB$, respectively.
    Let $K$, $L$, and $M$ be the midpoints of the segments $BP$, $CQ$, and $PQ$, respectively, and let $\Gamma$ be the circle passing through $K$, $L$, and $M$.
    Suppose that the line $PQ$ is tangent to the circle $\Gamma$.
    Prove that $OP = OQ$.
    \strut
  }
\end{equation}
This problem introduces ten named geometric objects and has a single goal.

Note that this class of problems does not admit a mathematical description but rather is defined empirically (\ie as those problems selected for olympiads).
The overwhelming majority of these problems are of a particular type -- plane geometry problems that can be expressed as problems in nonlinear real arithmetic (NRA).
However, while NRA is technically decidable, olympiad problems tend to be littered with order constraints and complex constructions (\eg mixtilinear incenter) and be well beyond the capability of existing algebraic methods.
On the other hand, they are selected to admit elegant, human-comprehensible proofs.
It is this class of problems for which the GMBL was designed to express;
though rare, any particular olympiad geometry problem is not guaranteed to be of this type and therefore is not necessarily expressible in the GMBL.

\subsection{Challenge: Globally Coupled Constraints}\label{sec:global}
A na\"{i}ve approach to generate models would incrementally instantiate objects via their immediate constraints.
For \eqref{eq:imo-2009-p2}, this would work as follows:
\begin{enumerate}
\item Sample points $A$, $B$, and $C$.
\item Compute $O$ as the circumcenter of $\Delta ABC$.
\item Sample $P$ and $Q$ on the segments $CA$ and $AB$, respectively.
\item Compute $K$, $L$, and $M$ as the midpoints of $BP$, $CQ$, and $PQ$, respectively.
\item Compute $\Gamma$ as the circle defined by $K$, $L$, and $M$.
\end{enumerate}
Immediately we see a problem -- there is no guarantee that $PQ$ is tangent to $\Gamma$ in the final model.
Indeed, the constraints of \eqref{eq:imo-2009-p2} are quite globally coupled -- the choice of $P$ partially determines the circle $\Gamma$
to which $PQ$ must be tangent, and every choice of $\Delta ABC$ does not even admit a pair $P$ and $Q$ satisfying this constraint.
This is an example of the frequent \emph{non-constructive} nature of IMO geometry problems.
When there is no obvious reparameterization
to avoid downstream effects, all constraints must be considered \emph{simultaneously} rather than incrementally or as a set of smaller local optimization problems.

\subsection{Challenge: Root Resolution}
\label{sec:root-resolution}

Even in the constructive case, local optimization is not necessarily sufficient given that multiple solutions can exist for algebraic constraints.
More specifically, two circles or a circle and a line intersect at up to two distinct points and in a problem that specifies each distinct intersection point, the correct root to assign is generally not locally deducible.
Without global information, this can lead to poor initializations becoming trapped in local minima.
The GMBL accounts for this by including a set of explicit \emph{root selectors} as described in Section \ref{sec:funcs-and-preds}.
These root selectors provide global information for selecting the appropriate point from a set of multiple solutions to a system of equations.
The prototypical example of this challenge is presented in Appendix \ref{appendix:local-info}.

\section{Methods}
\label{sec:methods}

In this section we present the GMBL and GMB in detail.
In our presentation, we make use of the following notation and definitions:
\begin{itemize}
\item
  The \codeword{type} of a geometric object can be one of (1) \codeword{point}, (2) \codeword{line}, or (3) \codeword{circle}.
  We denote the \codeword{type} of a real-valued number as \codeword{number}.
    \item We use \codeword{<>} to denote an instance of a type. 
    \item
      A \codeword{name} is a string value that refers to a geometric object.

\end{itemize}

\subsection{GMBL: Overview}

The GMBL is a DSL for expressing olympiad-level geometry problems that losslessly induces a numerical optimization problem.
It consists of four \emph{commands}, each of which has a direct interpretation regarding the accumulation of (1) real-valued parameters and (2) differentiable losses in terms of these parameters:

\begin{enumerate}
    \item \codeword{param}: assigns a \codeword{name} to a new geometric object parameterized either by a default or optionally supplied parameterization
    \item \codeword{define}: assigns a \codeword{name} to an object computed in terms of existing ones
    \item \codeword{assert}: imposes an additional constraint (\ie differentiable loss value) 
    \item \codeword{eval}: evaluates a given constraint in the final model(s) 
\end{enumerate}
Table~\ref{cmd-table} provides a summary of their usage.
The GMBL includes an extensible library of \emph{functions} and \emph{predicates} with which commands are written. 
Notably, this library includes a notion of \emph{root selection} to explicitly resolve the selection of roots to systems of equations with multiple solutions.

\subsection{GMBL: Commands}

In the following, we describe in more detail the usage of each command and their roles in constructing a tractable numerical optimization problem.

\codewordbf{param} accepts as arguments a \codeword{string}, a \codeword{type}, and an \emph{optional} parameterization.
This introduces a geometric object that is parameterized either by the default parameterization for \codeword{<type>} or by the supplied method.
Each primitive geometric type has the following default parameterization:
\begin{itemize}
    \item \codeword{point}: parameterized by its x- and y-coordinates
    \item \codeword{line}: parameterized by two points that define the line
    \item \codeword{circle}: parameterized by its origin and radius
\end{itemize}
Optional parameterizations embody our method's use of \emph{reparameterization} to decrease the number of parameters and increase model diversity.
For example, consider a point \codeword{C} on the line $\overleftrightarrow{AB}$ that is subject to additional constraints.
Rather than optimizing over the x- and y-coordinates of \codeword{C}, we can express \codeword{C} in terms of a single value $z$ that scales \codeword{C}'s placement on the line $\overleftrightarrow{AB}$. 

In addition to the standard usage of \codeword{param} outlined above, the GMBL includes an important variant of this command to introduce sets of points that form triangles and polygons.
This variant accepts as arguments (1) a \emph{list} of point names, and (2) a \emph{required} parameterization (see Table \ref{cmd-table}).
This \emph{joint} parameterization of triangles and polygons further prevents degeneracy.
For example, to initialize a triangle $\Delta ABC$, we can sample the vertices from normal distributions with means at distinct thirds of the unit circle.
This method minimizes the sampling of triangles with extreme angle values, as well as allows for explicit control over the distribution of acute vs. obtuse triangles by adjusting the standard deviations.
Appendix \ref{appendix:params-and-rs} includes a list of all available parameterizations. 

\begin{table*}[t]
  \begin{center}
    \caption{An overview of usage for the four commands.}
    \fontsize{10}{9}\selectfont
    \begin{tabular}{cc}
    \hline
    \textbf{Command} & \textbf{Usage} \\ \hline
    \codeword{param}            & \begin{tabular}[c]{@{}c@{}}\codeword{(param <string> <type> <optional-parameterization>)} \\ or \\ \codeword{(param (<string>, ..., <string>) <parameterization>)}\end{tabular} \\ \hline
    \codeword{define}  & \codeword{(define <string> <type> <value>)} \\ \hline
    \codeword{assert}           & \codeword{(assert <predicate>)} \\ \hline
    \codeword{eval}          & \codeword{(eval <predicate>)}  \\ \hline
    \end{tabular}
    \label{cmd-table}
  \end{center}
\end{table*}

\codewordbf{define} accepts as arguments a \codeword{string}, a \codeword{type}, and a \codeword{value} that is one of \codeword{<point>}, \codeword{<line>}, or \codeword{<circle>}.
This command serves as a basic assignment operator and is useful for caching commonly used values.
The functions described in Section \ref{sec:funcs-and-preds} are used to construct \codeword{<value>} from existing geometric objects.

\codewordbf{assert} accepts a single \codeword{predicate} and imposes it as an additional constraint on the system.
This is achieved by translating the \codeword{predicate} to a set of algebraic values
and registering them as losses.
This command does not introduce any \emph{new} geometric objects and can only refer to those already introduced by \codeword{param} or \codeword{define}.
Notably, dimensional constraints
and negations are always enforced via \codeword{assert}.
Detail on supported predicates is presented in Section \ref{sec:funcs-and-preds}.

\codewordbf{eval}, like \codeword{assert}, accepts a single \codeword{predicate} and therefore does not introduce any new geometric objects.
However, unlike \codeword{assert}, the corresponding algebraic values are evaluated and returned with the final model rather than registered as losses and enforced via optimization.
This command is most useful for those interested in integrating the GMBL with theorem provers.

\subsection{GMBL: Functions and Predicates}
\label{sec:funcs-and-preds}

The second component of our DSL is a set of functions and predicates for constructing arguments to the commands outlined above.
Functions construct new geometric objects and numerical values whereas predicates describe relationships between them.
Our DSL includes high-level abstractions for common geometric concepts in olympiad geometry (\eg excircle, isotomic conjugate).

\begin{figure*}[t]

\sbox\twosubbox{%
  \resizebox{\dimexpr.95\textwidth}{!}{%
    \includegraphics[height=3cm]{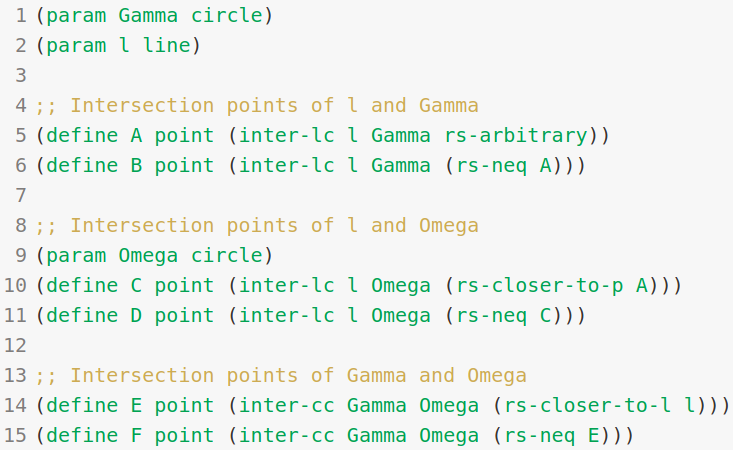}%
    \includegraphics[height=3cm]{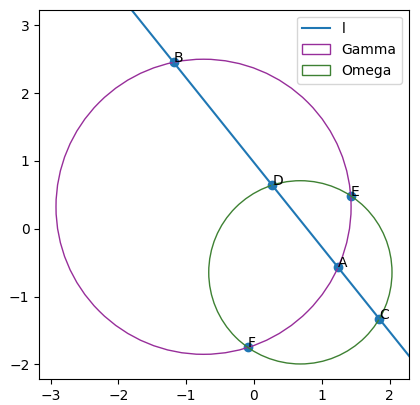}%
  }%
}
\setlength{\twosubht}{\ht\twosubbox}


\centering

\subcaptionbox{A GMBL program that uses root selectors.\label{fig:rs-program}}{%
  \includegraphics[height=\twosubht]{img/rs-program.png}%
}\quad
\subcaptionbox{A corresponding diagram.\label{fig:rs-fig}}{%
  \includegraphics[height=\twosubht]{img/rs-fig.png}%
}

\caption{An example usage of root selectors to resolve the intersections of lines and circles, and circles and circles.}
\label{fig:rs}
\end{figure*}

Functions in the GMBL employ a notion of \emph{root selectors} to address the ``multiple solutions problem'' described in Section \ref{sec:root-resolution}.
In plane geometry, this problem typically manifests with multiple candidate \codeword{point} solutions, such as the intersection between a line and a circle.
Root selectors control for this by allowing users to specify the appropriate \codeword{point} for functions with multiple solutions.
Figure \ref{fig:rs} demonstrates their usage in the functions \codeword{inter-lc} (intersection of a line and circle) and \codeword{inter-cc} (intersection of two circles).

Importantly, arguments to predicates and functions can be specified with functions rather than named geometric objects.
For a list of supported functions, predicates, and root selectors, refer to Appendices \ref{appendix:funcs}, \ref{appendix:preds}, and \ref{appendix:params-and-rs}, respectively.

\subsection{Auxiliary Losses}
\label{sec:aux-losses}

The optimization problem encoded by a GMBL progran includes three additional loss values.
Foremost, for every instance of a circle intersecting a line or other circle, we impose a loss value that ensures the two geometric objects indeed intersect.
The final two, albeit opposing losses are intended to minimize global degeneracy.
We impose one loss that minimizes the mean of all point norms to prevent exceptionally separate objects and a second to enforce a sufficient distance between points to maintain distinctness.

\subsection{Implementation}
\label{sec:optimization}

We built the GMB, an open-source implementation that compiles GMBL programs to optimization problems and generates models.
The GMB takes as input a GMBL program and processes each command in sequence to accumulate real-valued parameters and differentiable losses in a Tensorflow computation graph.
After registering auxiliary losses
, we apply off-the-shelf gradient-based local optimization to produce models of the constraint system.
In summary, to generate $N$ numerical models, our optimization procedure works as follows:
\begin{enumerate}
    \item Construct computation graph by sequentially processing commands.
    \item Register auxiliary losses. 
    \item Sample sets of initial parameter values and rank via loss value.
    \item Choose (next) best initialization and optimize via gradient descent.
    \item Repeat (4) until obtaining $N$ models or the maximum \# of tries is reached.
\end{enumerate}
Our program accepts as arguments (1) the \# of models desired ($\mathrm{default} = 1$), (2) the \# of initializations to sample ($\mathrm{default} = 10$), and (3) the max \# of optimization tries ($\mathrm{default} = 3$).
Our program also accepts the standard suite of parameters for training a Tensorflow model, including an initial learning rate ($\mathrm{default} = 0.1$), a decay rate ($\mathrm{default} = 0.7$), the max \# of iterations ($\mathrm{default} = 5000$), and an epsilon value ($\mathrm{default} = 0.001$) to determine stopping criteria.

\begin{table*}[t]
  \caption{
    An evaluation of our method's ability to generate a single model for each of the 36 IMO problems encoded in our DSL.
    For each problem, 10 sets of initial parameters were sampled over which our program optimized up to three.
    All data shown are the average of three trials.
    The first row demonstrates results using default parameters ($\epsilon = 0.001$, $\mathrm{learning\:rate} = 0.1$, $\mathrm{\#\:iterations} = 5,000$).}

  \begin{center}

    \fontsize{8}{8}\selectfont
    \begin{tabular}{ccccc*{3}{>{\centering\arraybackslash}m{3.75em}}}
\toprule
\multirow{2}{*}{$\boldsymbol{\epsilon}$} & \multirow{2}{*}{\textbf{\begin{tabular}[c]{@{}c@{}}Learning\\Rate\end{tabular}}} & \multirow{2}{2cm}{\centering\textbf{Iterations}} & \multirow{2}{*}{\textbf{\% Success}} & \multirow{2}{2cm}{\centering\textbf{\begin{tabular}[c]{@{}c@{}}\% Goal\\ Satisfaction\end{tabular}}} & \multicolumn{3}{c}{\centering\textbf{Time per Problem (s)}}  \\ \cmidrule(lr){6-8}
                                                  &  & &                                    &                                                                                                                  & \textbf{All}  & \textbf{Fail}   & \textbf{Success} \\ \midrule
0.001 & 0.1 & 5,000   & 93.52   & 85.84   & 27.07 & 223.51  & 14.72    \\
0.01 & 0.1 & 5,000   & 92.60      & 84.71   & 26.86  & 229.71  &  14.43    \\
0.001 & 0.01 & 5,000    & 88.88       &    86.32      & 27.54   & 137.85   & 14.33    \\
0.001 & 0.1 & 10,000    & 92.59     & 86.02      & 34.78  & 287.51 & 15.43    \\ \bottomrule
\end{tabular}

\end{center}
\label{tab:results}
\end{table*}

\section{Results}

In this section, we present an evaluation of our method's proficiency in three areas of expressing and solving olympiad-level geometry problems:
\begin{enumerate}
\item Expressing olympiad-level geometry problems as GMBL programs.
\item Generating models for these programs.
\item
  Preserving truths (up to tolerance) that are not directly optimized for.
\end{enumerate}
Table \ref{tab:results} contains a summary of our results.

Our evaluation considers all 39 IMO geometry problems since 2000.
Of these 39 problems, 36 can be expressed in our DSL.
Those that we cannot encode involve variable numbers of geometric objects.
For 32 of these 36 problems, we can express the goals as \codeword{eval} commands in the corresponding GMBL programs.
The goals of the additional four problems are not expressible in our DSL, \eg our DSL cannot express goals of the form ``Find all possible values of $\angle ABC$.''

To evaluate (2) and (3), we conducted three trials in which we ran our program on each of the 36 encodings with varying sets of arguments.
With default arguments, our program generated a single model for (on average) 94\% of these problems.
Our program ran for an average of 27.07 seconds for each problem but there is a stark difference between time to success and time to failure (14.72 vs 223.51 seconds) as failure entails completing all optimization attempts whereas successful generation of a model terminates the program.
We achieve similar success rates with more forgiving training arguments
or a higher tolerance. 

For use in automated theorem proving, it is essential that models generated by our tool not only satisfy the constraint problem up to tolerance but \emph{also} any other truths that follow from the set of input constraints.
The most immediate example of such a truth is the \emph{goal} of a problem statement.
Therefore, we used the goals of IMO geometry problems as a proxy for this ability by only checking the satisfaction of the goal in the final model (\ie with an \codeword{eval} statement) rather than directly optimizing for it.
In our experiments, we considered such a goal satisfied if it held up to $\epsilon * 10$ as it is reasonable to expect slightly higher floating-point error without explicit optimization.
Using default parameters, the goal held up to tolerance in 86\% of problems for which we found a model and could express the goal.
This rate was similar across all other sets of arguments.

\section{Future Work}

Here we discuss various opportunities for improvement of our method.

Firstly, improvements could be made to our method of numerical optimization.
While Tensorflow offers a convenient way of caching terms via a static computation graph and optimizing directly over this representation, there is not explicit support for \emph{constrained} optimization.
Because of this, arbitrary weights have to be assigned to each loss value.
Though rare, this can result in false positives and negatives for the satisfaction of a constraint.
Using an explicit constrained-optimization method (\eg SLSQP) would enable the separation of soft constraints (\eg maximizing the distance between points) and hard constraints (\eg those enforced by \codeword{assert}), removing the need for arbitrary weights.


Secondly, cognitive overhead could be reduced as users are currently required to determine degrees of freedom;
it would be far easier to write problem statements using only declarations of geometric objects and constraints between them, \eg using only \codeword{assert}.
This could be accomplished by treating our DSL as a low-level ``instruction set'' to which a higher-level language could be compiled.
The main challenge of such a compiler would be appropriately identifying opportunities to reduce the degrees of freedom. 
To achieve this, the compiler would require a decision procedure for line and circle membership.

Lastly, we could improve our current treatment of distinctness.
To prevent degenerate solutions, our method optimizes for object distinctness and rejects models with duplicates.
However, there is the occasional problem for which a local optimum encodes two provably distinct points as equal up to floating point tolerance.
There are many techniques that could be applied to this problem (\eg annealing) though we do not consider them here as the issue is rare.






\section{Related Work}
\label{sec:related-work}

Though many techniques for mechanized geometry diagram construction have been introduced over the decades, no method, to the best of our knowledge, can produce models for more than a negligible fraction of olympiad problems.
There exist many systems, built primarily for educational purposes, for interactively generating diagrams using ruler-and-compass constructions, \eg GCLC~\cite{janivcic2006gclc}, GeoGebra~\cite{hohenwarter2002geogebra}, Geometer's Sketchpad~\cite{scher1999lifting}, and Cinderella~\cite{richter2012cinderella}.
There are also non-interactive methods for deriving such constructions, \eg GeoView~\cite{bertot2004visualizing} and program synthesis~\cite{itzhaky2013solving, gulwani2011synthesizing}. 
However, as discussed in Section \ref{sec:global}, very few olympiad problems can be described in such a form. 
Alternatively, Penrose is an early-stage system for translating mathematical descriptions to diagrams that relies on constrained numerical optimization and therefore does not suffer from this expressivity limitation~\cite{ye2020penrose}.
However, this system lacks support for constraints with multiple roots, \eg intersecting circles.
There are more classical methods that similarly depart from constructive geometry.
MMP/Geometer~\cite{gao2002mmp} translates the problem to a set of algebraic equations and uses numerical optimization (\eg BFGS) and GEOTHER~\cite{wang2003automated, wang2002geother} first translates a predicate specification into polynomial equations, decomposes this system into representative triangular sets, and obtains solutions for each set numerically.
Neither of these programs are available to evaluate though we did test similar approaches using modern libraries (specifically: \codeword{sympy}~\cite{meurer2017sympy} and \codeword{scipy}~\cite{virtanen2020scipy}) and
both numerical and symbolic methods would almost always timeout on relatively simple olympiad problems.
Generating models for systems of geometric constraints is also a challenge in computer-aided design (CAD) for engineering diagram drawing.
Recent efforts focus on graph-based synthetic methods, a subset of techniques concerned with ruler-and-compass constructions~\cite{bettig2011geometric, hoffmann2002parametric, freeman1990incremental, kramer1990solving, latham1996connectivity, owen1991algebraic, fudos1995constraint}.
Most relevant to our method are Bettig and Shah's ``solution selectors'' which, similar to root selectors in the GMBL, allow users to specify the configuration of a CAD model~\cite{bettig2003solution}.
However, these solution selectors are purpose-built and do not generalize.

\section{Conclusion}

It is standard in GTP to rely on diagrams for heuristic reasoning but the scale of automatic diagram construction is limited.
To enable efforts to build a solver for IMO geometry problems, we developed a method for building diagrams for olympiad-level geometry problems.
Our method is based on the GMBL, a DSL for expressing geometry problems that induces (usually) tractable numerical optimization problems.
The GMBL includes a set of commands that have a direct interpretation for accumulating real-valued parameters and differentiable losses.
Arguments to these commands are constructed with a library of functions and predicates that includes notions of root selection, joint distributions, and reparameterizations to minimize degeneracy and the number of parameters.
We implemented our approach in an open-source tool that translates GMBL programs to diagrams.
Using this program, we evaluated our method on all IMO geometry problems since 2000.
Our implementation reliably produces models; moreover, known truths that are not directly optimized for typically hold up to tolerance.
By handling configurations of this complexity,
our system clears a roadblock in GTP and provides a critical tool for undertakers of the IMO Grand Challenge.

\bibliographystyle{abbrv}
\bibliography{main}

\clearpage

\appendix

\rowcolors{2}{white}{gray!15}

\section{Functions}
\label{appendix:funcs}

Here we provide a complete list of all functions supported in the GMBL.
Throughout this section, we provide examples that make use of the following declared geometric objects:
\begin{itemize}
    \item \codeword{A}, \codeword{B}, \codeword{C}, \codeword{D}, \codeword{E}, \codeword{F}, \codeword{G}, \codeword{H}, and \codeword{P} are instances of \codeword{point}
    \item \codeword{L1}, \codeword{L2}, and \codeword{L3} are instances of \codeword{line}
    \item \codeword{C1} and \codeword{C2} are instances of \codeword{circle}
    \item \codeword{N1} and \codeword{N2} are instances of \codeword{number}
\end{itemize}

\begin{longtable}{>{\centering\arraybackslash}m{6cm}>{\centering\arraybackslash}m{1.5cm}>{\centering\arraybackslash}m{4.5cm}}

\hline
\textbf{Function}                         & \textbf{Type} & \textbf{Description}                                                                      \\ \hline
\codeword{(amidp-opp A B C)}    & \codeword{point}         & The midpoint of the arc $\arc{AB}$ on $(ABC)$ excluding \codeword{C} \\
\codeword{(amidp-same A B C)}    & \codeword{point}         & The midpoint of the arc $\arc{ACB}$ on $(ABC)$\\
\codeword{(centroid A B C)}    & \codeword{point}         & The centroid of $\Delta ABC$ \\
\codeword{(circumcenter A B C)}    & \codeword{point}         & The circumcenter of $\Delta ABC$ \\
\codeword{(excenter A B C)}    & \codeword{point}         & The A-excenter of $\Delta ABC$ \\
\codeword{(foot A L1)}    & \codeword{point}         & The perpendicular foot from \codeword{A} to \codeword{L1} \\
\codeword{(harmonic-conj C A B)}    & \codeword{point}         & The harmonic conjugate of \codeword{C} w.r.t. $\overline{AB}$ \\
\codeword{(incenter A B C)}    & \codeword{point}         & The incenter of $\Delta ABC$ \\
\codeword{(inter-cc C1 C2 <root-selector>)} & \codeword{point}         & One of the intersections of \codeword{C1} and \codeword{C2}, as specified by a root selector          \\
\codeword{(inter-ll L1 L2)}                   & \codeword{point}         & The intersection of 2 lines                                                               \\
\codeword{(inter-lc L1 C1 <root-selector>)} & \codeword{point}         & One of the intersections of \codeword{L1} and \codeword{C1}, as specified by a root selector          \\
\codeword{(isogonal-conj D A B C)}                   & \codeword{point}         & The isogonal conjugate of \codeword{D} w.r.t. $\Delta ABC$       \\
\codeword{(isotomic-conj D A B C)}                   & \codeword{point}         & The isotomic conjugate of \codeword{D} w.r.t. $\Delta ABC$       \\
\codeword{(midp A B)}    & \codeword{point}         & The midpoint of $\overline{AB}$ \\
\codeword{(mixtilinear-incenter A B C)}    & \codeword{point}         & The A-mixtilinear incenter of $\Delta ABC$ \\
\codeword{(orthocenter A B C)}    & \codeword{point}         & The orthocenter of $\Delta ABC$ \\ \hline
\codeword{(connecting A B)}              & \codeword{line}          & The line connecting \codeword{A} and \codeword{B} (\ie $\overleftrightarrow{AB}$)                                                              \\
\codeword{(isogonal D A B C)}                 & \codeword{line}          & The isogonal of $\overleftrightarrow{AD}$ w.r.t. $\Delta ABC$                                \\
\codeword{(isotomic D A B C)}                 & \codeword{line}          & The isotomic of $\overleftrightarrow{AD}$ w.r.t. $\Delta ABC$                                \\
\codeword{(perp-bis A B)}                 & \codeword{line}          & The perpendicular bisector of $\overline{AB}$                                \\
\codeword{(perp-at A L1)}          & \codeword{line}          & The line through \codeword{A} that is perpendicular to \codeword{L1} \\ \hline
\codeword{(c3 A B C)}              & \codeword{circle}        & The circle passing through \codeword{A}, \codeword{B}, and \codeword{C}                                                            \\
\codeword{(circumcircle A B C)}        & \codeword{circle}        & The circumcircle of $\Delta ABC$                                             \\
\codeword{(excircle A B C)}    & \codeword{circle}         & The A-excircle of $\Delta ABC$ \\
\codeword{(incircle A B C)}        & \codeword{circle}        & The incenter of $\Delta ABC$                                             \\
\codeword{(mixtilinear-incircle A B C)}    & \codeword{circle}         & The A-mixtilinear incircle of $\Delta ABC$ \\
\codeword{(diam A B)}                    & \codeword{circle}        & The circle with diameter $\overline{AB}$                     \\ \hline
\codeword{(add N1 N2)}                    & \codeword{number}        & $\codeword{N1} + \codeword{N2}$       \\
\codeword{(area A B C)}                    & \codeword{number}        & The area of $\Delta ABC$       \\
\codeword{(dist A B)}                    & \codeword{number}        & The distance between \codeword{A} and \codeword{B}                                                            \\
\codeword{(div N1 N2)}                    & \codeword{number}        & $\codeword{N1} / \codeword{N2}$                                                            \\
\codeword{(mul N1 N2)}                    & \codeword{number}        & $\codeword{N1} * \codeword{N2}$                                                              \\
\codeword{pi}                    & \codeword{number}        & $\pi$    \\
\codeword{(pow N1 N2)}                    & \codeword{number}        & $\codeword{N1}^{\codeword{N2}}$    \\
\codeword{(neg N1)}                    & \codeword{number}        & $ - \codeword{N1}$                                                              \\
\codeword{(radius C1)}                    & \codeword{number}        & The radius of \codeword{C1}                                                              \\
\codeword{(sqrt N1)}                    & \codeword{number}        & $ \sqrt{\codeword{N1}}$                                                              \\
\codeword{(uangle A B C)}          & \codeword{number}        & The undirected angle formed by \codeword{A}, \codeword{B} and \codeword{C} (\ie $\angle A B C$)                                                   \\ \hline
\end{longtable}

\clearpage

\section{Predicates}
\label{appendix:preds}

Here we provide a complete list of all predicates supported in the GMBL.
The same variables are used as in Appendix \ref{appendix:funcs}:

\rowcolors{2}{white}{gray!15}

\begin{longtable}{>{\centering\arraybackslash}m{5.0cm}>{\centering\arraybackslash}m{7cm}}


\hline
\textbf{Predicate}                         & \textbf{Description}                                                                      \\ \hline
\codeword{(centroid P A B C)}           & \textbf{True} i.f.f. \codeword{P} is the centroid of $\Delta ABC$ \\
\codeword{(concur L1 L2 L3)}           & \textbf{True} i.f.f. \codeword{L1}, \codeword{L2}, and \codeword{L3} intersect at a single point \\
\codeword{(circumcenter P A B C)}           & \textbf{True} i.f.f. \codeword{P} is the circumcenter of $\Delta ABC$ \\
\codeword{(cong A B C D)}           & \textbf{True} i.f.f. $|\overline{AB}| = |\overline{CD}|$ \\
\codeword{(contri A B C D E F)}           & \textbf{True} i.f.f. $\Delta ABC \cong \Delta DEF$ \\
\codeword{(coll A B C)}          & \textbf{True} i.f.f. \codeword{A}, \codeword{B}, and \codeword{C} are collinear \\
\codeword{(cycl} $\codeword{P}_1$ \codeword{...} $\codeword{P}_N$\codeword{)}          & \textbf{True} i.f.f. $\codeword{P}_1$ \codeword{...} $\codeword{P}_N$ lie on a common circle, where $\codeword{P}_1$ \codeword{...} $\codeword{P}_N$ are points and $N \geq 4$ \\
\codeword{(= A B)}          & \textbf{True} i.f.f. $\codeword{A} = \codeword{B}$ \\
\codeword{(= N1 N2)}          & \textbf{True} i.f.f. $\codeword{N1} = \codeword{N2}$ \\
\codeword{(eq-ratio A B C D E F G H)}           & \textbf{True} i.f.f. $\frac{|\overline{AB}|}{|\overline{CD}|} = \frac{|\overline{EF}|}{|\overline{GH}|}$ \\
\codeword{(foot P A L1)}         & \textbf{True} i.f.f. \codeword{P} is the perpendicular foot from \codeword{A} to \codeword{L1} \\
\codeword{(> N1 N2)}          & \textbf{True} i.f.f. $\codeword{N1} > \codeword{N2}$ \\
\codeword{(>= N1 N2)}          & \textbf{True} i.f.f. $\codeword{N1} \geq \codeword{N2}$ \\
\codeword{(incenter P A B C)}           & \textbf{True} i.f.f. \codeword{P} is the incenter of $\Delta ABC$ \\
\codeword{(inter-ll P L1 L2)}           & \textbf{True} i.f.f. \codeword{P} is the intersection of \codeword{L1} and \codeword{L2} \\
\codeword{(< N1 N2)}          & \textbf{True} i.f.f. $\codeword{N1} < \codeword{N2}$ \\
\codeword{(<= N1 N2)}          & \textbf{True} i.f.f. $\codeword{N1} \leq \codeword{N2}$ \\
\codeword{(midp P A B)}         & \textbf{True} i.f.f. \codeword{P} is the midpoint of $\overline{AB}$ \\
\codeword{(on-circ P C1)}           & \textbf{True} i.f.f. \codeword{P} is on \codeword{C1} \\
\codeword{(on-line P L1)}           & \textbf{True} i.f.f. \codeword{P} is on \codeword{L1} \\
\codeword{(on-ray P A B)}           & \textbf{True} i.f.f. \codeword{P} is on $\overrightarrow{AB}$ \\
\codeword{(on-seg P A B)}           & \textbf{True} i.f.f. \codeword{P} is on $\overline{AB}$ \\
\codeword{(opp-sides A B L1)}          & \textbf{True} i.f.f. \codeword{A} and \codeword{B} are on opposite sides of \codeword{L1} \\
\codeword{(orthocenter P A B C)}           & \textbf{True} i.f.f. \codeword{P} is the orthocenter of $\Delta ABC$ \\
\codeword{(perp L1 L2)}          & \textbf{True} i.f.f. $L1 \perp L2$ \\
\codeword{(para L1 L2)}          & \textbf{True} i.f.f. $L1 \parallel L2$ \\
\codeword{(same-side A B L1)}          & \textbf{True} i.f.f. \codeword{A} and \codeword{B} are on the same side of \codeword{L1} \\
\codeword{(sim-tri A B C D E F)}           & \textbf{True} i.f.f. $\Delta ABC \sim \Delta DEF$ \\
\codeword{(tangent-cc C1 C2)}           & \textbf{True} i.f.f. \codeword{C1} is tangent to \codeword{C2} \\
\codeword{(tangent-lc L1 C1)}           & \textbf{True} i.f.f. \codeword{L1} is tangent to \codeword{C1} \\
\codeword{(tangent-at-cc A C1 C2)}           & \textbf{True} i.f.f. \codeword{C1} is tangent to \codeword{C2} at \codeword{A }\\
\codeword{(tangent-at-lc A L1 C1)}           & \textbf{True} i.f.f. \codeword{L1} is tangent to \codeword{C1} at \codeword{A }\\
\hline
\end{longtable}

\clearpage

\section{Parameterizations and Root Selectors}
\label{appendix:params-and-rs}

Here we provide an enumeration of all parameterizations and root selectors supported in our DSL. The same variables are used as in Appendix \ref{appendix:funcs}.

The following table provides a complete list of all \emph{optional} parameterizations for standard usage of \codeword{param}.
The value in the \textbf{Type} column denotes the \codeword{type} for which a parameterization is used:

\rowcolors{2}{white}{gray!15}

\begin{table*}[h]
\begin{center}
\begin{tabular}{>{\centering\arraybackslash}m{5cm}>{\centering\arraybackslash}m{1.5cm}>{\centering\arraybackslash}m{5.5cm}}
\hline
\textbf{Parameterization}                         & \textbf{Type} & \textbf{Description}                                                                      \\ \hline
\codeword{(on-circ C1)}    & \codeword{point}         & Parameterizes a point on the circle \codeword{C1} \\
\codeword{(on-line L1)}    & \codeword{point}         & Parameterizes a point on the line \codeword{L1} \\
\codeword{(on-major-arc C1 A B)}    & \codeword{point}         & Parameterizes a point on the major arc of \codeword{C1} connecting \codeword{A} and \codeword{B}. \\
\codeword{(on-minor-arc C1 A B)}    & \codeword{point}         & Parameterizes a point on the minor arc of \codeword{C1} connecting \codeword{A} and \codeword{B}. \\
\codeword{(in-poly} $\codeword{P}_1$ \codeword{...} $\codeword{P}_N$\codeword{)}      & \codeword{point}    & Parameterizes a point inside the polygon with vertices $\codeword{P}_1$ \codeword{...} $\codeword{P}_N$, where $\codeword{P}_1$ \codeword{...} $\codeword{P}_N$ are points and $N \geq 3$ \\
\codeword{(on-ray A B)}    & \codeword{point}         & Parameterizes a point on the ray $\overrightarrow{AB}$ \\
\codeword{(on-ray-opp A B)}    & \codeword{point}         & Parameterizes a point beyond \codeword{A} on the ray $\overrightarrow{BA}$ \\
\codeword{(on-seg A B)}    & \codeword{point}         & Parameterizes a point on the segment connecting \codeword{A} and \codeword{B} \\

\hline
\codeword{(tangent-lc C1)}    & \codeword{line}         & Parameterizes a line tangent to \codeword{C1}\\
\codeword{(through A)}    & \codeword{line}         & Parameterizes a line passing through \codeword{A}\\ \hline
\codeword{(tangent-cc C1)}    & \codeword{circle}         & Parameterizes a circle tangent to \codeword{C1}\\
\codeword{(tangent-cl L1)}    & \codeword{circle}         & Parameterizes a circle tangent to \codeword{L1}\\
\codeword{(through A)}    & \codeword{circle}         & Parameterizes a circle passing through \codeword{A}\\
\codeword{(origin A)}    & \codeword{circle}         & Parameterizes a circle centered at \codeword{A}\\
\codeword{(radius N1)}    & \codeword{circle}         & Parameterizes a circle with radius \codeword{N1}\\
\hline
\end{tabular}
\end{center}
\end{table*}

Alternatively, the following table provides a complete list of all supported parameterizations for introducing \emph{sets} of points as triangles and polygons with \codeword{param}. Note that type information is omitted, as these parameterizations are always used to introduce lists of type  \codeword{(point, ..., point)}. Examples are provided as usages are subtle:

\newpage

\rowcolors{2}{white}{gray!15}

\begin{table*}[h]
\begin{center}
\begin{tabular}{>{\centering\arraybackslash}m{3.75cm}>{\centering\arraybackslash}m{5cm}>{\centering\arraybackslash}m{3cm}}
\hline
\textbf{Parameterization}           & \textbf{Example}              & \textbf{Meaning}                                                                      \\ \hline
\codeword{acute-tri}                                                                        & \codeword{(param (P Q R) acute-tri)}    & \codeword{P}, \codeword{Q}, and \codeword{R} are parameterized as an acute triangle     \\
\codeword{(acute-iso-tri <name>)}  & \begin{tabular}[c]{@{}l@{}}\codeword{(param (P Q R)} \\\ \ \ \ \codeword{(acute-iso-tri Q))}\end{tabular}         & \codeword{P}, \codeword{Q}, and \codeword{R} are parameterized as an acute isoscles triangle with $QP = QR$         \\
\codeword{(iso-tri <name>)}                                        & \begin{tabular}[c]{@{}l@{}}\codeword{(param (P Q R)} \\\ \ \ \ \codeword{(iso-tri Q))}\end{tabular}              & \codeword{P}, \codeword{Q}, and \codeword{R} are parameterized as an isoscles triangle with $QP = QR$           \\
\codeword{(right-tri <name>)}                                    & \begin{tabular}[c]{@{}l@{}}\codeword{(param (P Q R)} \\\ \ \ \ \codeword{(right-tri Q))}\end{tabular} & \codeword{P}, \codeword{Q}, and \codeword{R} are parameterized as a right triangle with a right angle at $\angle PQR$ \\
\codeword{triangle}                                    & \codeword{(param (P Q R) triangle)}            & \codeword{P}, \codeword{Q}, and \codeword{R} are parameterized as a triangle \\
\codeword{polygon}                                                                         & \codeword{(param (P Q R S) polygon)}               & \codeword{P}, \codeword{Q}, \codeword{R}, and \codeword{S} are parameterized to form a convex quadrilateral      \\
\hline
\end{tabular}
\end{center}
\end{table*}

Lastly, the following table provides a complete list of all root selectors:

\rowcolors{2}{white}{gray!15}

\begin{longtable}{ccc}

\hline
\textbf{Root Selector}                         & \textbf{Description}                                                                      \\ \hline
\codeword{rs-arbitrary}          & An arbitrary root \\
\codeword{(rs-neq A)}    & A root that is not equal \codeword{A} \\
\codeword{(rs-opp-sides A L1)}    & A root on the opposite side of \codeword{L1} as \codeword{A} \\
\codeword{(rs-same-side A L1)}    & A root on the same side of \codeword{L1} as \codeword{A} \\
\codeword{(rs-closer-to-p A)}    & The root that is closer to \codeword{A} \\
\codeword{(rs-closer-to-l L1)}    & The root that is closer to \codeword{L1} \\
\hline
\end{longtable}

\clearpage

\section{Challenge: Local vs. Global Optimization}
\label{appendix:local-info}

Here we present a distilled example where global optimization is required to reliably produce models.
Optimization is tasked with assigning $A$ and $B$ to the two distinct intersection points of a line and a circle.
\begin{figure*}[h]
  \centering
  \includegraphics[height=6cm,keepaspectratio]{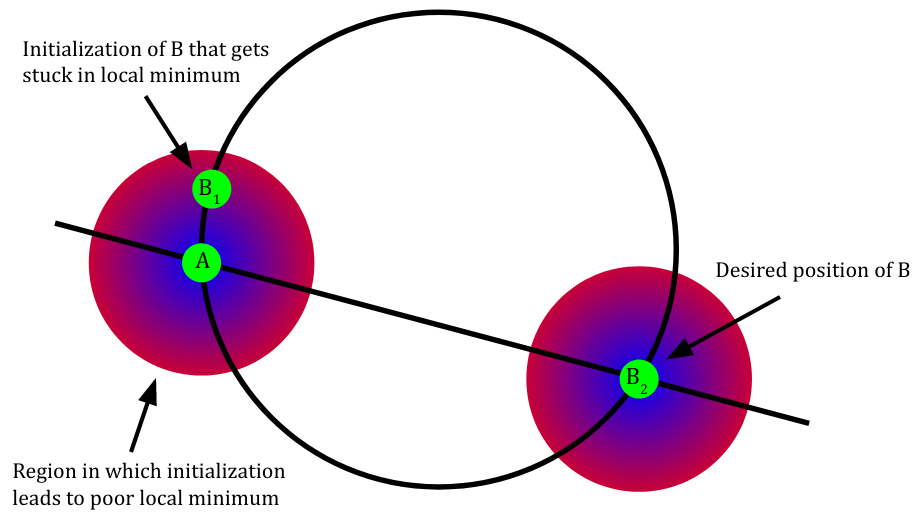}
\end{figure*}
To do so, the distance from each point from both the line and circle will be minimized while maintaining a minimum distance between points.
$A$ will be arbitrarily assigned to one of the two choices.
However, without global knowledge of the root occupied by $A$, an initialization of $B$ too close to $A$ will be trapped in a local minimum.
Blue indicates a region of low loss and red a region of high loss.
Points are represented as green circles.

\clearpage

\section{Examples}
\label{appendix:examples}

Here we provide a gallery of example diagrams generated by our tool and their associated problem statements (both in text and as GMBL programs).

\subsection*{IMO 2011, Problem 6}

\begin{center}
\emph{IMO Problem Statement}
\begin{tcolorbox}[width=12cm,colback=gray!10,arc=0pt,auto outer arc]
  Let $ABC$ be an acute triangle with circumcircle $\Gamma$.
  Let $\ell$ be a tangent line to $\Gamma$, and let $\ell_a$, $\ell_b$ and $\ell_c$ be the lines obtained by reflecting $\ell$ in the lines $BC$, $CA$, and $AB$, respectively.
  Show that the circumcircle of the triangle determined by the lines $\ell_a$, $\ell_b$ and $\ell_c$ is tangent to the circle $\Gamma$.

\end{tcolorbox}
\end{center}

\begin{center}
\emph{GMBL Problem Statement}
\begin{tcolorbox}[width=12cm,colback=gray!10,arc=0pt,auto outer arc]

  \codeword{(param (A B C) acute-tri)}

  \codeword{(define Gamma circle (circumcircle A B C))}

  \codeword{(param l line (tangent-lc Gamma))}

  \codeword{(define la line (reflect-ll l (line B C)))}

  \codeword{(define lb line (reflect-ll l (line C A)))}

  \codeword{(define lc line (reflect-ll l (line A B)))}

  \codeword{(eval (tangent-cc Gamma (circumcircle (inter-ll la lb)}

  \hphantom{Micros}\codeword{(inter-ll la lc) (inter-ll lb lc))))}

\end{tcolorbox}
\end{center}

\begin{center}
\emph{Diagram}
\\
\includegraphics[height=8cm,keepaspectratio]{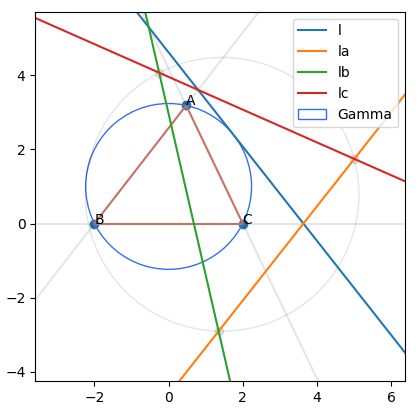}
\end{center}

\newpage

\subsection*{IMO 2008, Problem 1}

\begin{center}
\emph{IMO Problem Statement}
\begin{tcolorbox}[width=12cm,colback=gray!10,arc=0pt,auto outer arc]
  An acute-angled triangle $ABC$ has orthocentre $H$.
  The circle passing through $H$ with centre the midpoint of $BC$ intersects the line $BC$ at $A_1$ and $A_2$.
  Similarly, the circle passing through $H$ with centre the midpoint of $CA$ intersects the line $CA$ at $B_1$ and $B_2$, and the circle passing through $H$ with centre the midpoint of $AB$ intersects the line $AB$ at $C_1$ and $C_2$.
  Show that $A_1$, $A_2$, $B_1$, $B_2$, $C_1$, $C_2$ lie on a circle.
\end{tcolorbox}
\end{center}

\begin{center}
\emph{GMBL Problem Statement}
\begin{tcolorbox}[width=12cm,colback=gray!10,arc=0pt,auto outer arc]
  \fontsize{8.5}{11}\selectfont

  \codeword{(param (A B C) acute-tri)}

  \codeword{(define H point (orthocenter A B C))}
  \\

  \codeword{(define A1 point (inter-lc (line B C) (coa (midp B C) H) rs-arbitrary))}


  \codeword{(define A2 point (inter-lc (line B C) (coa (midp B C) H) (rs-neq A1)))}

  \codeword{(define B1 point (inter-lc (line C A) (coa (midp C A) H) rs-arbitrary))}


  \codeword{(define B2 point (inter-lc (line C A) (coa (midp C A) H) (rs-neq B1)))}

  \codeword{(define C1 point (inter-lc (line A B) (coa (midp A B) H) rs-arbitrary))}


  \codeword{(define C2 point (inter-lc (line A B) (coa (midp A B) H) (rs-neq C1)))}
  \\

  \codeword{(eval (cycl A1 A2 B1 B2 C1 C2))}
\end{tcolorbox}
\end{center}

\begin{center}
\emph{Diagram}
\\
\includegraphics[height=8cm,keepaspectratio]{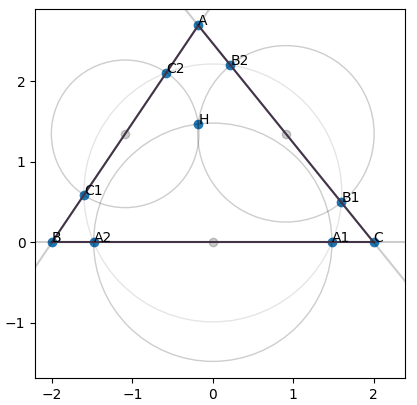}
\end{center}

\newpage

\newpage

\subsection*{IMO 2009, Problem 2}

\begin{center}
\emph{IMO Problem Statement}
\begin{tcolorbox}[width=12cm,colback=gray!10,arc=0pt,auto outer arc]
  Let $ABC$ be a triangle with circumcentre $O$.
  The points $P$ and $Q$ are interior points of the sides $CA$ and $AB$, respectively.
  Let $K$, $L$, and $M$ be the midpoints of the segments $BP$, $CQ$, and $PQ$, respectively, and let $\Gamma$ be the circle passing through $K$, $L$, and $M$.
  Suppose that the line $PQ$ is tangent to the circle $\Gamma$.
  Prove that $OP = OQ$.
\end{tcolorbox}
\end{center}

\begin{center}
\emph{GMBL Problem Statement}
\begin{tcolorbox}[width=12cm,colback=gray!10,arc=0pt,auto outer arc]

  \codeword{(param (A B C) triangle)}

  \codeword{(define O point (circumcenter A B C))}
  \\
  \codeword{(param P point (on-seg C A))}

  \codeword{(param Q point (on-seg A B))}
  \\
  \codeword{(define K point (midp B P))}

  \codeword{(define L point (midp C Q))}

  \codeword{(define M point (midp P Q))}

  \codeword{(define Gamma circle (circ K L M))}
  \\
  \codeword{(assert (tangent-lc (line P Q) Gamma))}

  \codeword{(eval (cong O P O Q))}
\end{tcolorbox}
\end{center}

\begin{center}
\emph{Diagram}
\\
\includegraphics[height=8cm,keepaspectratio]{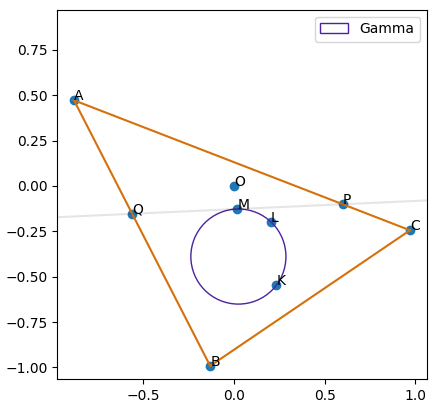}
\end{center}

\end{document}